# AlGaAs/GaAs/AlGaAs quantum wells as a sensitive tool for the MOVPE reactor environment

V. Dimastrodonato, L.O. Mereni, R. J. Young[1] and E. Pelucchi

*Tyndall National Institute, University College Cork, Cork, Ireland*

Corresponding author: e-mail valeria.dimas@tyndall.ie, Phone: +353 21 420 4195, Fax: +353 21 4270271, address: Tyndall National Institute "Lee Maltings", Prospect Row, Cork, Ireland

## Abstract

**We present in this work a simple Quantum Well (QW) structure consisting of GaAs wells with AlGaAs barriers as a probe for measuring the performance of arsine purifiers within a MetalOrganic Vapour Phase Epitaxy system. Comparisons between two different commercially available purifiers are based on the analysis of low temperature photoluminescence emission spectra from thick QWs, grown on GaAs substrates misoriented slightly from (100). Neutral excitons emitted from these structures show extremely narrow linewidths, comparable to those which can be obtained by Molecular Beam Epitaxy in an ultra-high vacuum environment, suggesting that purifications well below the 1ppb level are needed to achieve high quality quantum well growth.**

---

[1] Department of Physics, Lancaster University, Lancaster, LA1 4WY, UK

## 1. Introduction

Three-five arsenide (and phosphide) alloys are at the very core of many everyday-use and industrial semiconductor devices, such as lasers, optical modulators and high electron mobility transistors. Such components have become the main building blocks in a wide range of electronic applications, including optical storage data systems, mobile telephones, radio telescopes, satellite television receivers and navigation devices [1]. The successful employment of such components relies on both good quality epitaxial layers and cost effective, large production scale facilities. When epitaxial structures are needed in the context of industrial applications and device production, Metalorganic Vapour Phase Epitaxy (MOVPE) is often the preferred crystal growth technique. It indeed has some significant advantages, as generally recognized, over other production methods, such as Molecular Beam Epitaxy (MBE): short system downtimes, maintainability, long term reproducibility, scalability to multiwafer growth, relatively "easy" operators training.

III-V MOVPE grown nanostructures are also becoming increasingly attractive in a number of intriguing fundamental physics topics which were, till a few years ago, a prerogative of MBE grown materials. Quantum wires and dots for quantum cavity electrodynamics experiments, quantum information technologies and fundamental physics [2,3], quantum cascade lasers and polariton cavities [4,5] or confined nanostructures for probing transport effects in the quantum regime [6] can now be grown by MOVPE. In the last twenty years in fact a general progress of the reactor technologies, an enhancement of source material quality, together with a substantial development in metalorganic and hydride purification techniques, led to a noteworthy improvement of all MOVPE outputs [7].

Although "MBE standard" structures can be nowadays grown by MOVPE by more and more groups worldwide, a large disparity in the material quality obtained by different research communities remains. An indication, generally speaking, that a basic understanding on some of the key parameters, which need to be controlled and diagnosed to guarantee the best results, is still lacking. This is also one of the issues we address in our manuscript.

Different approaches have been proposed in the literature to monitor impurity incorporation during MOVPE growth with the aim of improving the reactor overall performances. Typical examples include growing thick GaAs [8] or AlGaAs layers [9,10] and a subsequent analysis of the intensity of their optical properties. Previous works focused on the emission properties of low temperature near band edge, where donors and acceptors features can often be recognized. GaAs/AlGaAs QW or 2D electron gas systems, acting as active layers in most devices and/or physical structures, were also presented as efficient tool to monitor the impact of impurity incorporation on the device performances [10,11,12]. No reference in the literature of what should be the best route towards material optimization can though be found, also because different applications might involve "differently" optimized material.

Despite the general necessity of employing sources with lower impurity levels [8], a profound analysis on the real impact that modern purification processes have on the optical (and transport) properties of III-V compounds and devices is still lacking. Different commercial purifiers in fact claim different purification performances, but the exact quantitative impact of small water and Oxygen impurities (~1 ppb or less) on the properties of III-V QWs or 2D electron gases grown by MOVPE has not been thoroughly analyzed yet.

While it is not the objective of this paper to compare the different methods utilized in the literature to assess the reactor impurity levels, our manuscript specifically addresses the potential of using GaAs QWs grown on slightly misoriented GaAs wafers – recently demonstrated as capable of "MBE like quality" [13] – as a very efficient and selective monitor of the reactor environment. Reference 13 by Pelucchi et al. is the only report up to now which has overcome the historical divide between MBE and MOVPE on this particular material system (AlGaAs barriers are known to be particularly keen on Oxygen and Carbon incorporation): in general MOVPE has proven unable of delivering QW excitonic linewidths below the few meV level, while thick GaAs QWs (~25 nm) grown by MBE could achieve linewidths as low as ~0.3 meV [14].

A number of variables affect the photoluminescence linewidth at cryogenic temperatures of a QW. As schematically depicted in Fig. 1, the excitonic wavefunction is interacting with a number of features present into a QW and its barriers, which in turn affect the long and short range *potential template* experienced by the exciton, and so its emission energy: interface roughness, alloy disorder in the barrier (an alloy is never a perfectly random structure, resulting in a local modulation of the alloy composition), and the presence of impurities in the barrier and QW materials [15,16,17]. Other parameters/variables not shown in Fig.1 also play a role, e.g. the spectral meandering induced by the local electric fields associated with free charges, the coupling to phonons and the carriers relaxation times if a perfect thermalization is not allowed [15,16,17,18].

In general thick QWs are the less sensitive to interface roughness, and conversely predicted/found to be, at cryogenic temperatures, extremely responsive to unintentional impurities (which can be very low in MBE), due to the aforementioned exciton interaction with the effective "atomic-scale roughness" associated with impurity centre [14, 15,16,17,18].

We introduce in this work an analysis of the relevance of hydride purification for the achievement of "*MBE like*" quality in this fundamental material system, and explicitly discuss a technological approach for a reproducible path towards the achievement of high material quality, which can be employed in any existing MOVPE laboratory. In particular, we tested the two commercially available arsine purifiers, claiming the highest purification specifications at the time of the experiments, to analyze for differences in their performances [ 19 ]. For comparison sake, we intentionally grew the same identical structure (i.e. all parameters with the exception of the arsine purity were the same). Record low temperature QW linewidths were obtained with both purifiers and, notably, the "activation/conditioning" process of these purifiers is verified as a crucial step that can critically delay (as much as one year) the attainment of the best purification levels.

## 2. Experimental procedure

*System description*. All our samples were nominally identical and grown in an Aixtron 200 1x2" horizontal reactor. The system is equipped with purification systems for the carrier gas (Nitrogen) and the group V sources (Arsine and Phosphine). The Nitrogen was purified with a heated purifier (SAES Getters Monotorr PS4MT15) and the hydride lines are equipped with a double dry purification system. The system allows selection of which of (either/or/both) the two purifiers are to be used for a determined growth run (see Fig 2 for a schematic of the purification panel). The Nitrogen and hydride impurity levels were monitored with ceramic moisture sensors (Michell Pura), located downstream the purifiers and capable of detecting ~1ppb of $H_2O$. The purification set-up for each hydride line was designed to allow the two purifiers to be put in series if required, and is equipped with a "purging" line (served by the same purified Nitrogen of the MOVPE system) which enables a Nitrogen purging of the purifier lines up to the "vent-run" switching valves (not shown in Fig. 2). We tested two newly installed Arsine purifiers: a Mykrolis CE450KF-SK4B (from now on referred to as AP1) and a SAES Getters MicroTorr PG1-502FV-151 (AP2). Both purifiers are *nominally* capable of purification performances below 1ppb of impurities (for $H_2O$ and $O_2$). The reactor is equipped with an Ebara dry pump and the sample introduction occurs via a 7N closed cycle Nitrogen glove box to avoid air exposure of the reactor body.

*Experimental conditions*. All samples characterized for this manuscript were grown at 20 mbar (to keep the partial pressure of unintentional impurities as low as possible) using Trimethylgallium, Trimethylaluminium (both Cambridge Chemical Company) and highest purity grade Arsine as sources [20]. All growths were performed following full conditioning of the purifiers as from manufacturer's instructions [21]. All lines were Nitrogen-purged until the dew point detectors were at the minimum reading of the full scale (<< 1ppb of $H_2O$). Following reference 13, for simplicity and consistency with our standard laboratory practice and previously reported results, all samples included 3 GaAs quantum wells (2, 5 and 15nm thick) separated by 300nm thick $Al_{0.3}Ga_{0.7}As$ barriers, with a 30nm thick GaAs cap. The buffer consisted of 500 nm GaAs and 500 nm $Al_{0.3}Ga_{0.7}As$ layers. The quoted layer thicknesses and concentrations are the nominal values. The growth rate was maintained at 1 μm/hour (~1 monolayer/s) for the $Al_{0.30}Ga_{0.70}As$ layers. The growth temperature (estimated surface temperature) was 690°C, a V/III ratio of 230 in the AlGaAs barriers and 130 in the GaAs active layer was chosen, and no growth interruption was performed. The substrate was misoriented by 0.2 degree off ± 0.02 [13] towards (111)A from (100).

Particular attention was also paid to the reactor environment quality and to temperature control: all the samples were grown with reactor's inner surface baked and coated by material from several processes. Photoluminescence (PL) spectra were acquired at ~10K with the samples mounted on the cold finger of a micro-photoluminescence closed cycle cryostat. We excited our samples with a laser emitting at 1.88 eV (658 nm). All the PL spectra reported in this manuscript have been obtained in a "defocused" micro-PL set-up, with a spot size of ~5 micron diameter and an excitation level of a few tens microW, optimizing excitation form measurement to measurement for the narrowest linewidth possible. As reported in Ref 13 or 14 the charged exciton intensity is known to have a strong dependence on the excitation conditions and sample temperature (2 Kelvin can make a significant difference on the charged exciton PL intensity). In this paper we concentrate on the free exciton features, and will not repeat the (interesting but well established) discussion on the charged exciton behaviour [13, 14, 16]. It should be also said that we checked in all our samples for localization effects and compared macroPL conditions (few mm spot size) versus microPL conditions (1 micron spotsize) without detecting any significant spectral difference, consistently with what reported in Ref. 13. This is indeed indicative of a strong uniformity in the growth process, and a lack of significant disorder at our interfaces. All Full Width at Half Maximum (FWHM) values reported here are extracted from a Lorentzian fit of the correspondent peak.

### 3. Results and discussion

Ref 13 shows that it is possible to obtain a low-temperature PL linewidth FWHM as low as ~0.65 meV (and low-temperature electron mobilities greater than $10^6$ cm$^2$/Vs) in thick GaAs/Al$_{0.30}$Ga$_{0.70}$As QWs produced by MOVPE on slightly misoriented GaAs

(001) substrates (< 1 degree), while thinner QWs are more sensitive to the interface roughness and show broader FWHM. For this reason we will concentrate through our manuscript only on the 15 nm QW, without discussing the results on the thinner QWs present in our samples. The 2 and 5 nm QWs have only been used as a reference to check that our samples showed/behaved similarly to Ref 13, guarantying us from possible artefacts arising from unintentional process mistakes.

The small misorientation modifies the surface morphology of the GaAs substrate in a significant and controllable manner, with optimal QW optical properties observed on surfaces that exhibit the onset of the transition from step flow to step bunching, a feature peculiar of the MOVPE process and which has been associated with the surface diffusion and decomposition of the metalorganic precursors [ 22 ]. The resulting interfaces for the thick QWs, in the optimum case, show little influence on the photoluminescence linewidths, which strongly depend, on the other hand, on the unintentional background doping [14,23,24].

There is in fact a striking dependence of the QW optical properties on extremely small lattice misorientation variations. A 15 nm QW grown in the same growth run on separate GaAs substrates, with intentional misorientations in the range zero to one degree towards the (111)A planes, shows a surprising variety of low-temperature PL linewidths, ranging from a few meV (< 5) to less than 1 meV, with maxima at the two extreme of the interval and an optimum minimum (FWHM ~0.6 meV) in the 0.15-0.3° off (100) range [13]. It must also be noted that there is only a narrow window of growth parameters where such optimal linewidths can be obtained, and particular attention has to be paid on the growth temperature and V/III ratios, which need to be different in the AlGaAs and GaAs layers. It is then important to carefully calibrate the reactor temperature if the results reported in Ref. 13 and in this manuscript have to be reproduced.

A natural extension to this discovery is the possibility, within the other options, to use the spectral purity of a QW emission to monitor the unintentional impurity level background in a reactor. The Arsine purification quality was soon found to be an enabling key parameter and we investigated specifically for differences in the performances of Arsine commercial purifiers. We particularly addressed this problem by growing always *nominally identical samples*, so that any oscillations of the FWHM values could be directly related to the impurity levels, while the interface roughness and surface morphology would stay constant [25].

While Nitrogen (and Hydrogen or Argon) purification levels can be monitored by means of special mass spectrometry down to a few hundreds of ppt (for water and Oxygen), for hydride sources this possibility is still lacking [8, 26]. In general, commercial providers cannot certify purification levels below 1ppb (especially when water and Oxygen are considered), notwithstanding the fact that they all informally present better performances of their purifier, despite the technical difficulties in demonstrating it. This poses the question: which of the commercially available purifiers is best suited for obtaining the highest level of material quality?

In Fig 3 we show the photoluminescence spectrum from a 15 nm thick GaAs QW in AlGaAs barriers, grown on a 0.2 degree off substrate and obtained a few months after the installation of our reactor with AP1. The growth was performed after AP1 was fully conditioned, and the Arsine lines were subsequently purged under Nitrogen to remove the substantial $H_2O$ contamination induced by the conditioning procedure [21]. Several calibration runs were subsequently performed, the reactor was fully "conditioned" and covered with GaAs before the QW test runs were performed.

Fig. 3 illustrates a dominant charged exciton peak (the assignment has been based on the extensive studies reported, for example, in references 13, 14 and 27) while the neutral exciton peak shows a weak intensity with a FWHM of 3.7 meV, in line with the results reported by most MOVPE groups [8, 28]. While this growth run is not to be considered as a poor quality QW by MOVPE standards (it indeed is as good as most of the best reports to date), it certainly does not show the less than 1 meV FWHM of the record value reported in the literature [13]. It has to be stressed that this result is not due to a badly controlled growth process, or to an out of specification wafer choice. Atomic Force Microscopy (AFM) was used to measure the actual misorientation of the wafer and we always found a good agreement between nominal specifications and the actual wafer misorientation, confirming that the GaAs cap step dynamics on the surface of our QW samples was either step flow or at the onset of step bunching (i.e. with a mixed behaviour) as in Ref. 13. An example of the typical stepped surface is shown in Fig. 4: the flattened AFM image (the flattening was performed in the direction perpendicular to the step orientation) presents only step flow surface morphology. The lacking evidence of a step bunching behaviour confirms a low value of the actual misorientation, whose simple calculation is illustrated in the inset.

As the performance of AP1 was unsatisfactory and did not initially appear to be improving with time, we proceeded to test the other commercially available purifier. Fig 5 a) shows photoluminescence from one of the first 15 nm QWs grown six months after the installation of the reactor and a few weeks after the installation of the AP2 purifier with the hydride line fully recovered from the contamination following the conditioning procedure. In this case the neutral linewidth has a reduced FWHM of ~1 meV as from our fitting procedure. The purification process continued to improve in the following weeks with the Arsine line usage, and a few weeks later a QW with record FWHM (Fig 5 b) for this reactor was achieved, with a neutral exciton linewidth of just 0.6 meV.

It is important at this point to clarify the reasons why in this report we are neglecting to discuss the AlGaAs barriers optical properties. AlGaAs features are generally difficult to detect in our samples, as thermalization towards the QWs is a very efficient process, and the AlGaAs signal is present only at relatively high pump intensities, where spectral features tend to be broadened and not significant for our purposes anymore. Moreover, in our previous studies, the AlGaAs excitonic features did not appear to show such striking dependence on the reactor quality (the general trend is the same, but less accentuated, so we did not rely on that for our studies) [29]. This could be understood as the QWs represent the minimum of the potential energy in the system and actually act as sink for all charges associated to impurities, effectively multiplying the QWs sensitivity to the impurity concentration while depleting the bulk. The limited analysis conducted on our samples did not show significant deviations from what was reported in Ref. 13 and 29.

The Arsine quality continued to improve using both purifiers in series (first AP1 and then AP2): we were able to achieve the best photoluminescence linewidth for a GaAs/AlGaAs quantum well to date. The FWHM of the neutral exciton shown in Fig. 6a) is ~412 μeV (as extracted by the Lorentzian fitting procedure, which we stress has no ambiguity in its outputs), a value which represents an absolute record for MOVPE growths and is very close to the best results of ~300 μeV achieved by MBE. It should be observed that the charged exciton peak is not a sign of poor sample quality, as it is always present even in the best MBE grown samples [13,14,27,30].

Following a large number of growths and significant usage, the performance of AP1 improved. Fig 6 b) shows the emission spectrum of a 15nm thick QW, whose growth, under the same conditions as all other samples, was carried out 15 months after system installation (and more than 10 Kg of Arsine flown trough the purifier) with only AP1, AP2 being in bypass mode. The FWHM is very narrow, and close to the best result achieved, indicating that AP1 features are similar to AP2 after a sufficient purification time.

The results obtained here suggest that the best commercially available purifiers offer very good performance, but a significant running-in period is necessary before reaching purification levels guaranteeing high quality sample growth. This period of "activation" was just a few weeks for AP2, but more than a year for AP1. Both achieved their specified <1ppb level of purification within a few weeks of their recommended conditioning procedure (resulting in around 1ppm of $H_2O$ contamination of the hydrides lines), as monitored by our dew point sensor, after a Nitrogen purging of the lines. Nevertheless, a level of ~1ppb $H_2O$ does not appear sufficient to achieve <1 meV FWHM in these QW samples. Perhaps the purification grade should be comparable to that assured by the Nitrogen purifiers (~200-300 ppt) in order to obtain the record QW FWHM demonstrated here. Notably both AP2 and AP1 provided the excellent results, despite the substantially different activation periods. This difference is probably a result of the way the purifiers are packaged and "activated" at the factory level.

It is important to observe that our results are not affected by impurities absorbed by the hydride lines after the "conditioning" procedures or subsequent growth runs. The hydride lines were constantly purged with purified Nitrogen for long periods of time to ensure no such effects were present. The purification performances are then to be considered as the result of the purifier assembly alone, without memory effects induced by the stainless steel pipes and valves.

We would also like to stress that the approach we present here with our QWs should not be in anyway considered as the only and exclusive tool to assess reactor impurities and understand which parameters need to be improved. It is well known that bulk (GaAs/AlGaAs etc) spectroscopy helps in determining to some extent the origin of the impurities, and extensive work has been done in the past (see references of our introduction) to analyse their effects. There is no doubt that more extensive and demanding investigations (in terms of time and resources) like photoluminescence excitation spectroscopy or absorption measurements and the like would nicely complement our results, helping to quantitatively assess the quality of the MOVPE sources and purification techniques. On the other hand our QW method appears to be simple and effective in determining reactor quality to a level which guarantees a high quality of throughputs. In fact, we underline that our careful reactor handling and monitoring have produced high quality samples in a broad range of applications in our laboratory: for example alongside the results presented with this work it has also allowed growth of some of the best (highest spectral purity) InGaAs-based site-controlled quantum dots [31], further supporting the importance of a correct purification strategy. Ref. 31 reports record excitonic linewidths, as narrow as 18 $\mu$eV under non resonant pumping, which represented a substantial improvement over all MOVPE grown QDs reported to date [7].

## 4. Conclusions

In conclusion, commercial Arsine purifiers with the highest specifications are capable of providing excellent purification performance, allowing for near-MBE quality III-V material samples to be grown by MOVPE. The record linewidth FWHM from thick (15 nm) GaAs/AlGaAs QWs here presented is significantly lower than the previous best reported value by MOVPE and is close to the best values obtained by MBE. Hydride purifiers do not instantaneously perform at their best, but require a period of "activation" through constant usage before an impurity level which is less than 1ppb can be attained. Different commercially available purifiers can require very different "activation" times, suggesting that the "factory" purifier treatment and packaging are of great importance when trying to achieve the best results.

## 5. Acknowledgments

This research was enabled by the Irish Higher Education Authority Program for Research in Third Level Institutions (2007-2011) via the INSPIRE programme, and by Science Foundation Ireland under grants 05/IN.1/I25 and 07/SRC/I1173. We are grateful to K. Thomas for his support with the MOVPE system. We also would like to acknowledge the contribution of G. Knight to the photoluminescence data acquisition.

V. Dimastrodonato et al.

**References**

[1] See, for example: Special Issue on High-Speed Circuits: 2002 GaAs IC Symposium, in IEEE Journal of solid-state circuits, 38 (2003).

[2] K. A. Atlasov, K. F. Karlsson, E. Deichsel, A. Rudra, B. Dwir, and E. Kapon, Appl. Phys. Lett. 90 (2007) 153107; P. Gallo, M. Felici, B. Dwir, K. A. Atlasov, K. F. Karlsson, A. Rudra, A. Mohan, G. Biasiol, L. Sorba, and E. Kapon, Appl. Phys. Lett. 92 (2008) 263101.

[3] E. Pelucchi, S. Watanabe, K. Leifer, B. Dwir, Q. Zhu, P. De Los Rios, and E. Kapon, NanoLetters 7 (2007) 1282; Q. Zhu, K. F. Karlsson, E. Pelucchi, and E. Kapon, NanoLetters 7 (2007) 2227; M. H. Baier, A. Malko, E. Pelucchi, D. Oberli, D. Chek-al-kar, and E. Kapon, Phys. Rev. B 73 (2006) 205321.

[4] J. S. Roberts, R. P. Green, L. R. Wilson, E. A. Zibik, D. G. Revin, J. W. Cockburn, and R. J. Airey, Appl. Phys. Lett. 82 (2003) 4221; M. Troccoli, D. Bour, S. Corzine, G. Höfler, A. Tandon, D. Mars, D J. Smith, L. Diehl, and F. Capasso, Appl. Phys. Lett. 85 (2004) 5842; L. Sirigu, A. Rudra, E. Kapon, M. I. Amanti, G. Scalari, and J. Faist, Appl. Phys. Lett. 92 (2008) 181111.

[5] H. Cao, S. Pau, and Y. Yamamoto, G. Bjork, Phys Rev B 54 (1996) 8083; D. Sanvitto, D. N. Krizhanovskii, D. M. Whittaker, S. Ceccarelli, and M. S. Skolnick, J. S. Roberts, Phys. Rev. B 73 (2006) 241308.

[6] M. Merano, S. Sonderegger, A. Crottini, S. Collin, P. Renucci, E. Pelucchi, A. Malko, M. Baier, E. Kapon, B. Deveaud, and J-D. Ganière, Nature 438 (2005) 479; E. Levy, A. Tsukernik, M. Karpovski, A. Palevski, B. Dwir, E. Pelucchi, A. Rudra, E. Kapon, and Y. Oreg, Phys. Rev. Lett. 97 (2006) 196802; L. Meier, G. Salis, I. Shorubalko, E. Gini, S. Schön, and K. Ensslin, Nature Physics 3 (2007) 650-654.

[7] D. Richter, R.t Hafenbrak, K. D Jons, W-M. Schulz, M Eichfelder, MHeldmaier, R Roßbach, M Jetter and P. Michler, Nanotechnology 21 (2010) 125606.

[8 ] J. Feng, R. Clement, and M. Raynor, J. Crystal Growth 310 (2008) 4780.

[9 ] S. Leu, F. Hoehnsdorf, W. Stolz, R. Becker, A. Salzmann, A. Greiling, J. Crystal Growth 195 (1998) 98.

[10] H. C. Chui, B. E. Hammons, N. E. Harff, J. A. Simmons, and M. E. Sherwin, Appl. Phys. Lett. 68 (1996) 208.

[11] G. Bastard, C. Delalande, M. H. Meynadier, P. M. Frijlink, and M. Voos, Phys Rev. B 29 (1984) 7042.

[12] Jan P. van der Ziel, X. Tang, and R. Johnson, Appl. Phys. Lett. 71 791 (1997).

[13] E. Pelucchi, N. Moret, B. Dwir, D.Y. Oberli, A. Rudra, N. Gogneau, A. Kumar, E. Kapon, E. Levy and A. Palevski , J. Appl. Phys. 99 (2006) 093515.

[14] A. Esser, E. Runge, R. Zimmermann, and W. Langbein, Phys. Rev. B 62 (2000) 8232.

[15] F. Grosse and R. Zimmermann, Superlattices and Microstructures 17 (1995) 439; R. Zimmermann, F. Grosse and E. Runge, Pure & Appl . Chem., Vol. 69 (1997) 1179-1186.

[16] V. Savona and W. Langbein, Rev. B 74 (2006) 075311.

[17] C. Ropers, M. Wenderoth, L. Winking, T. C. G. Reusch, M. Erdmann, and R. G. Ulbrich, S. Malzer and G. H. Döhler, Phys. Rev. B 75 (2007) 115317.

[18] M. Gurioli, A. Vinattieri, J. Martinez-Pastor, and M. Colocci, Phys. Rev. B 50 (1994) 16.

[19] The other commercially available purifiers claim performances on the oxygen and water level purification that are > 50 times worse than the two purifiers we tested. It will be clear from our analysis that even 1ppb of water is critical in achieving narrow

linewidths in the AlGaAs/GaAs/AlGaAs QW system: as a consequence we did not invest into testing them, also in view of the substantial financial cost this would have implied. It should be said that technical papers on some of the existing commercial purifiers seem to suggest that their performances are better than the producer specified values (see Ref.8). Nevertheless it is not the subject of this paper to analyze the performances of all commercial purifiers available, but only that of underlining the importance of an appropriate purification strategy.

[20] Air Products MB3. Matheson also produces a similar grade arsine.

[21] Both AP1 and AP2 need to be "conditioned" before usage: as indicated by instruction manuals, a given total flow of Arsine needs to pass through the purifier for several hours before it can be operated. During the conditioning procedure, the purifier surfaces get saturated with Arsenic/Arsine and only after this "saturation" the purifiers are ready to work. Both purifiers while conditioning released significant amount of water (as measured with our ceramic moisture sensors), resulting in an approximate 1ppm (or slightly less) $H_2O$ contamination of the lines downstream the purifier.

[22] A. L.-S. Chua, E. Pelucchi, A. Rudra, B. Dwir, E. Kapon, A. Zangwill, and D. D. Vvedensky, Appl. Phys. Lett. 92 (2008) 013117.

[23] N. Moret, D.Y. Oberli, E. Pelucchi, N. Gogneau, A. Rudra and E. Kapon, Appl. Phys. Lett. 88 (2006) 141917.

[24] See also: A. Rudra, E. Pelucchi, D. Oberli, N. Moret, B. Dwir and E. Kapon , J. Crystal Growth 272 (2004) 615.

[25] All samples have been grown also with the same hydride and metallorganic sources, all far from "depletion" effects. We indeed stress that in our comparison the only variable which changed over time is the arsine purification levels.

[26] P. Giannoules, C. Solcia, and M. Succi, Semiconductor Fabtech, N. 2 (1995) 153-155; H. H. Funke, B. L. Grissom, C. E. McGrew, and M. W. Raynor, Rev.Sci.Inst., 74, 3909, 2003.

[27] G. Eytan, Y. Yayon, M. Rappaport, H. Shtrikman, and I. Bar-Joseph, Phys. Rev. Lett. 81, (1998),1666.

[28] See for example: R.D. Dupuis, J.G. Neff, and C.J. Pinzone, J. Crystal Growth 124 (1992) 558; M. Shinohara, H. Yokoyama and N. Inoue, J Vac Sci Technol B13 (1995) 1773.

[29] N. Moret, PhD thesis, NO 4070 (2008), ÉCOLE POLYTECHNIQUE FÉDÉRALE DE LAUSANNE.

[30] We are aware that, as only exception to our statement, at Sheffield University it was possible to achieve extraordinary QW linewidths as low as 150 microeV with no charged excitonic shoulder. Maxime Hugues, private communication.

[31] L.O. Mereni, V. Dimastrodonato, R.J. Young and E. Pelucchi, Appl. Phys. Lett. 94 (2009) 223121.

**Figure captions**

Fig.1: schematic representation of the effects of interface roughness, impurity incorporation (black symbols) and alloy disorder (oval symbols -different colours correspond to different concentrations-) on excitons confined in a GaAs QW embedded in between AlGaAs barriers.

Fig.2: Schematic of the purification panel. The two purifiers AP1 and AP2 are connected in series through a simple valves system. In this sketch the bypass valves (in red) are both closed and $AsH_3$ gas passes through both the purifiers. By closing $P_{IN}$ (purifier input) and $P_{OUT}$ (purifier output) valves of the purifier to be bypassed and opening the bypass valve (B), it is possible to use only one of the purifiers and test its performances. Note that the Nitrogen purging lines are not depicted in this figure.

Fig.3: μ-PhotoLuminescence emission spectrum at low temperature of a 15nm thick $GaAs/Al_{0.3}Ga_{0.7}As$ QW grown by purging $AsH_3$ precursor through AP1 purifier only, after its conditioning procedure. The charged ($X^-$) and weak neutral (X) excitons are indicated and the dominant charged exciton emission suggests the need of a purification status improvement.

Fig. 4: Atomic Force Microscopy image of a typical stepflow-like surface on a GaAs substrate (100) with a nominal miscut of ±0.2° off toward (111)A. In the inset the calculation of the actual misorientation θ is illustrated: the length L, equal to 1μm in this image, is measured from the AFM scan size, the height h is given by the number of the steps multiplied by their height (~1ML). The lack of a step bunching morphology corroborates the small value of the actual miscut angle.

Fig.5: Spectra of emission from two different QWs. Both the purifiers were employed during the growths and the runs were performed after conditioning the AP2 purifier. A remarkable improvement of the performances was reached after a few weeks: the FWHM of the neutral exciton peak narrows from 1meV (a) to 0.6meV (b). In fig. a) and b) Lorentzian fits of charged and neutral excitons are shown for a clear identification of the peaks and all FWHM values are extrapolated from their relative data.

Fig. 6: a) Emission peak from our best QW, grown with the two purifiers in series. A striking purity reached after a long purification time of $AsH_3$ gas turns out into a FWHM of the neutral exciton peak as narrow as 0.412meV. b) Emission spectrum of a QW grown by purifying $AsH_3$ through AP1 purifier only, bypassing AP2. A notable improvement of the purification level reached by the AP1 purifier was obtained (see Fig. 3 for comparison): the neutral exciton peak presents a FWHM very close to our best result. Note that the *total* Lorentzian fit perfectly matches with the charged exciton fit, therefore it's not clearly visible in the figure.

**Figure 1**

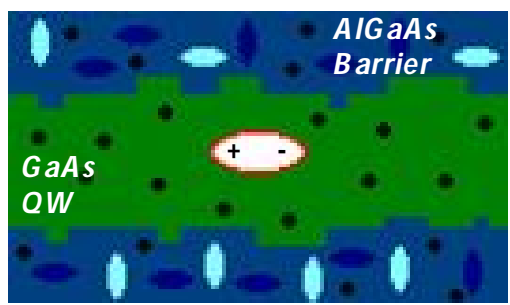

**Figure 2**

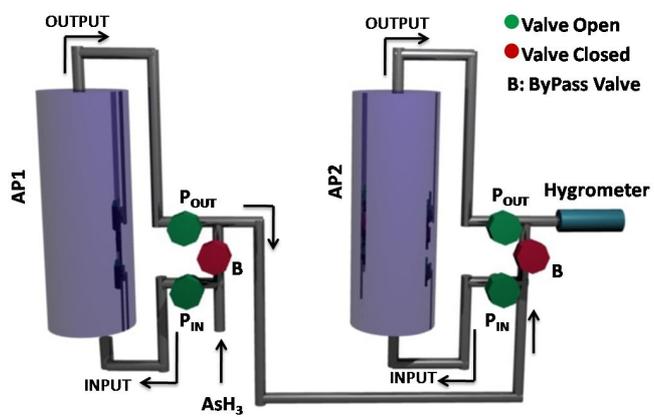

**Figure 3**

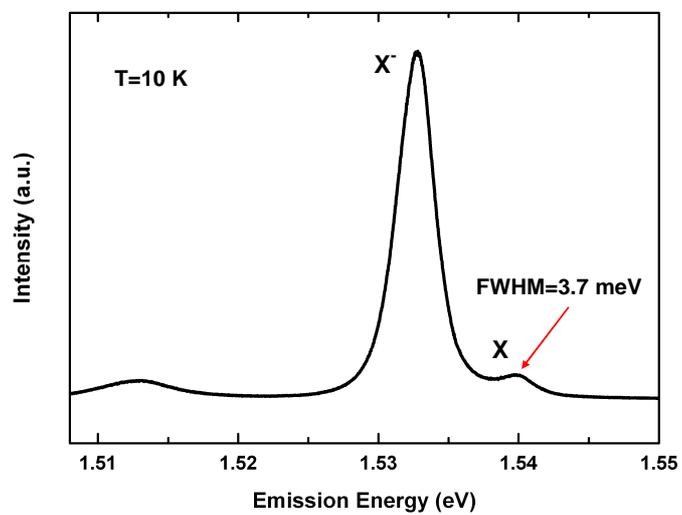

**Figure 4**

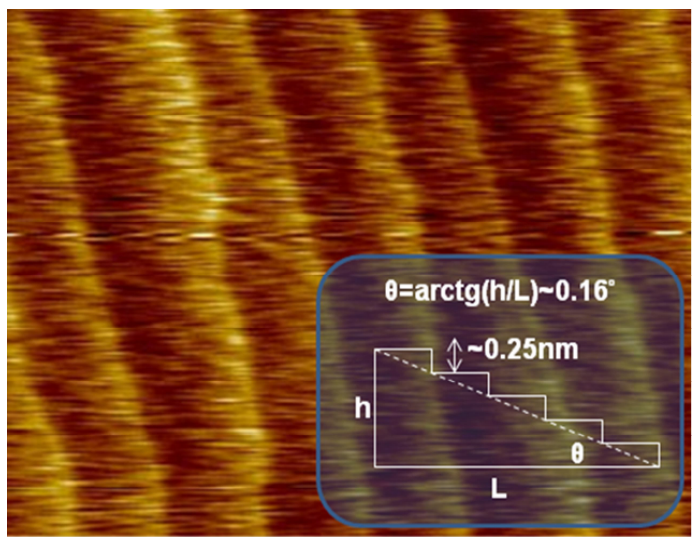

V. Dimastrodonato et al.

**Figure 5**

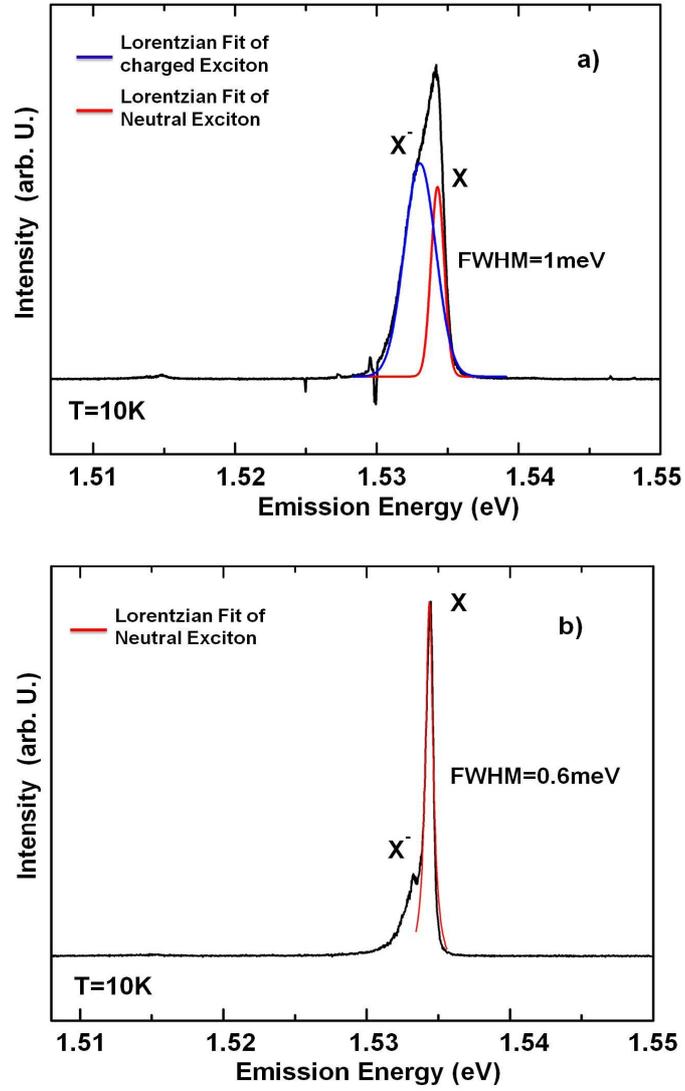

**Figure 6**

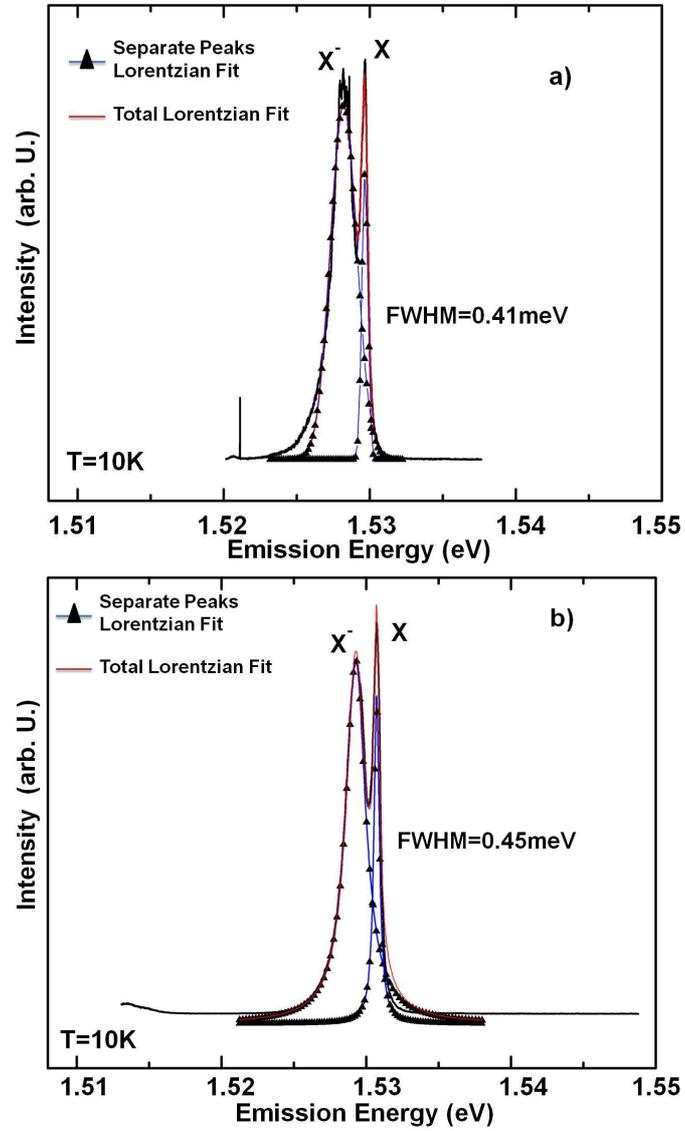